\documentclass[lettersize,journal]{IEEEtran}
\usepackage{graphicx} % Required for inserting images
\usepackage{amsmath}
\usepackage{bbm}
\usepackage{amssymb}
\usepackage{subcaption}
\usepackage{placeins}
\usepackage{balance}
\usepackage{url}
\makeatletter
\usepackage{xcolor}

\title{Quantifying Distribution Shift in Traffic Signal Control with Histogram-Based GEH Distance}
\author{
\IEEEauthorblockN{Federico Taschin\IEEEauthorrefmark{1}\IEEEauthorrefmark{2}} \and 
\IEEEauthorblockN{Ozan Tonguz\IEEEauthorrefmark{2}\IEEEauthorrefmark{3}} \\
\IEEEauthorblockA{\IEEEauthorrefmark{1}KTH Royal Institute of Technology}  \\ 
\IEEEauthorblockA{\IEEEauthorrefmark{2}Virtual Traffic Lights} \\
\IEEEauthorblockA{\IEEEauthorrefmark{3}Carnegie Mellon University}
}
\date{}

\begin{document}

\maketitle

\begin{abstract}
Traffic signal control algorithms are vulnerable to distribution shift, where performance degrades under traffic conditions that differ from those seen during design or training. This paper introduces a principled approach to quantify distribution shift by representing traffic scenarios as demand histograms and comparing them with a GEH-based distance function. The method is policy-independent, interpretable, and leverages a widely used traffic engineering statistic. We validate the approach on 20 simulated scenarios using both a NEMA actuated controller and a reinforcement learning controller (FRAP++). Results show that larger scenario distances consistently correspond to increased travel time and reduced throughput, with particularly strong explanatory power for learning-based control. Overall, this method can predict performance degradation under distribution shift better than previously published techniques. These findings highlight the utility of the proposed framework for benchmarking, training regime design, and monitoring in adaptive traffic signal control.  
\end{abstract}

\section{Introduction}
Traffic signal control is a cornerstone of urban mobility, with direct implications for congestion, safety, and sustainability. Modern approaches range from rule-based methods such as fixed-time and actuated control to adaptive systems and reinforcement learning (RL). Despite their differences, all these systems share a critical vulnerability: their performance can degrade when faced with traffic conditions that differ substantially from those observed during design, calibration, or training. This phenomenon, known in the machine learning literature as distribution shift, has only recently been recognized as a key challenge in traffic signal control.

In practice, distribution shift manifests itself when a traffic intersection experiences traffic demand patterns not well represented in the data used to configure the controller. Examples include changes in commuter behavior, seasonal variations, or non-recurrent events such as accidents or roadworks. While the effects of such changes are well documented—ranging from increased delay to severe congestion—there is currently no established methodology for quantifying how different two traffic scenarios are. Without such a measure, researchers and practitioners lack a systematic way to anticipate performance degradation, compare the robustness of control algorithms, or design training regimes that cover the relevant spectrum of conditions.

This paper addresses that gap by introducing a simple but effective representation of traffic scenarios and a distance measure to quantify scenario-to-scenario differences. The core idea is to represent each intersection’s traffic demand as a set of histograms over volumes on the intersection traffic movements. This representation captures the distribution of flows through the traffic intersection in a form that is both interpretable and comparable across scenarios.

Our main contributions are the following:
\begin{itemize}
    \item A method to characterize traffic scenarios by constructing histograms of vehicular demand over time for each traffic movement in a single-intersection setting.

    \item An extension of the GEH measure -widely used in traffic engineering- to histogram-based representations that allows it to be used for quantifying the difference between two traffic scenarios with high precision. 

    \item An empirical validation of the proposed method using both a NEMA-style actuated controller and an RL-based controller, demonstrating that the proposed distance measure positively correlates performance degradation of the traffic control algorithm with distribution shift.

    \item An empirical comparison of our method with other methods in the literature, demonstrating the superiority of this approach in the single-intersection case for both NEMA-style and RL controllers.
\end{itemize}
Our method works better than previously published techniques and immediately enables computing regression coefficients that allow predicting performance degradation when a traffic scenario changes. While previous methods only considered hourly volumes, our method can use fine-grained traffic flow data with minute-level precision, allowing to recognize short-lived events (such as short traffic spikes) that would otherwise be neglected by other methods. 

By providing improved methods to measure distribution shift, this work represents an important step toward predictive tools that can inform both the design and the deployment of traffic signal control systems. Beyond immediate applications in benchmarking and algorithm evaluation, the proposed metrics facilitate adaptive training regimes, scenario selection, and long-term monitoring of deployed systems under evolving traffic conditions.

\section{Related Work}

Distribution shift is a long-standing challenge in traffic signal control. Studies have already recognized that deviations from assumed traffic conditions can severely degrade the performance of fixed-time signal plans \cite{berry_1956, Papageorgiou_2003}. This phenomenon, commonly referred to as \textit{ageing}, arises when traffic patterns change over time and render preconfigured timings suboptimal \cite{stevanovic2006assessing, qadri_2020}. Short-term disturbances such as incidents or surges in demand further complicate operations. Actuated controllers provide limited adaptability by responding to detector inputs, but their performance deteriorates when actual conditions diverge from the calibration data \cite{shelby_2001}. Adaptive systems such as SCOOT \cite{scoot1990} and SCATS \cite{scats1980} mitigate ageing by continuous retuning, yet they remain insufficiently responsive to non-recurrent events \cite{taylor1998incident} and can underperform in oversaturated conditions \cite{stevanovic2019benefits}.  

Measuring differences between traffic scenarios has therefore been of interest in transportation engineering. The GEH statistic \cite{feldman2012geh} is widely used in model validation thanks to its scale-insensitive properties and accepted thresholds. Alternative measures include absolute flow differences across movements 
\cite{bell1986ageing} and robustness testing of signal timings in microscopic simulation \cite{stevanovic2006assessing}. Other approaches generate perturbed scenarios indirectly, for example by modifying origin–destination matrices \cite{Dutta2008EvaluationOT}, synthetically altering volumes \cite{park2001}, or sampling from empirical data \cite{Ostojic2017AssessmentOT}. While these studies provide insight into controller robustness, they do not provide a formal and interpretable scenario-to-scenario distance.  

In reinforcement learning for traffic signal control, generalization across traffic conditions has recently gained attention and traction. Metalight \cite{metalight} extends the FRAP architecture with meta-training to adapt more quickly to unseen scenarios, but does not provide a numerical measure of scenario difference. Generalight \cite{generalight} leverages generative modeling using the Wasserstein GAN framework \cite{wgan} to synthesize diverse scenarios, indirectly introducing a similarity notion via the Wasserstein distance. However, because it operates at the level of entire traffic routes, two scenarios that are functionally similar at the intersection (e.g., producing the same turning flows) may nonetheless be assigned large distances, leading to counterintuitive results. 

To the best of our knowledge, the only explicit attempt to quantify distribution shift in traffic scenarios are \cite{tonguz2025kl} and \cite{tonguz2025ks} which use Kullback–Leibler divergence \cite{kl} and Kolmogorov–Smirnov distance to measure the difference in traffic flows. However, these methods consider only the hourly vehicle volumes, neglecting the demand variations that can occur within the observed time period.

In summary, while distribution shift is widely acknowledged in both traffic engineering and RL-based traffic signal control, there is no robust, established, and interpretable metric for reliably quantifying the difference between traffic scenarios. This paper addresses this gap by introducing a histogram-based representation of demand combined with a GEH distance function, providing a practical and domain-relevant measure of distribution shift.

\section{Methodology}

\subsection{Traffic Scenario}
We begin by providing a definition of a \textit{traffic scenario} and how it can be represented in a way that allows numerical comparison with other traffic scenarios.

A traffic scenario is characterized by the road network topology together with the traffic demand traversing it. In this work, we constrain ourselves to a fixed topology of a single intersection, therefore, we will not consider the topology in the remainder of this paper. 

\begin{figure}
    \centering
    \includegraphics[width=\linewidth]{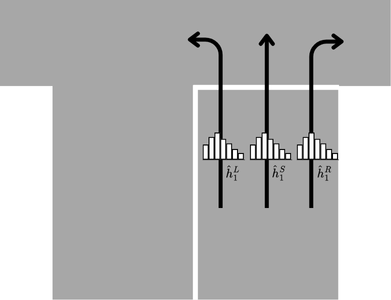}
    \caption{Representation of an approach of a traffic intersection. The approach has three allowed traffic movements, that vehicles choose depending on their intended turn direction. To each movement, we assign an histogram that represents the vehicle demand over time.}
    \label{fig:traffic_movements}
\end{figure}

Let $M$ denote the set of allowed traffic movements of the traffic intersection (see Figure \ref{fig:traffic_movements} for an example). For a given scenario $S$, the traffic demand can be summarized by the empirical distribution of vehicles observed over $M$. Intuitively, two scenarios may share the same total vehicle volume but differ significantly in how those vehicles are distributed across traffic movements (for example, predominantly straight versus predominantly turning movements). It is precisely these differences in distribution that often drive the effectiveness—or failure—of a traffic signal control strategy.

To capture this formally, we represent a scenario $S$ as a collection of histograms $\{h_m\}_{m\in M}$, where each histogram $h_m$ counts the number of vehicles assigned to traffic movement $m$ over the time period into consideration. 
%This representation has two desirable properties:
%\begin{itemize}
%    \item Comparability: two scenarios defined on the same network can be compared directly at the level of their histograms, without requiring knowledge of the specific routes taken by vehicles.
%    \item Interpretability – the histogram representation aligns naturally with how traffic engineers think about volumes on approaches and movements, making it easier to interpret differences.
%\end{itemize}

Figure \ref{fig:traffic_movements} shows three traffic movements of an intersection approach. To each movement, we assign an histogram that represents the vehicle demand over time for that movement.

Given two scenarios $S_1$ and $S_2$, the problem of measuring distribution shift reduces to computing a distance function $d(S_1, S_2)$ between their corresponding sets of histograms. 

\subsection{Histogram Construction}
Let the simulation horizon be $[0, T]$, we partition it into $K$ non-overlapping intervals of equal duration $\Delta = \frac{T}{K}$. For each traffic movement $m$ we define a histogram
\begin{equation}
    h_m = \big(h_m(1), h_m(2), \dots, h_m(K)\big)
\end{equation}
where $h_m(k)$ is the number of vehicles that reach the intersection for traffic movement $m$ in the time-interval $k$.

\subsubsection{Scenario representation}
A scenario $S$ is represented by the set of histograms representations, one for each traffic movement.
\begin{equation}
\mathcal{H}(S) = \{ h_m : m \in M \},
\end{equation}
Comparing two scenarios $S_1, S_2$ thus reduces to comparing their corresponding histogram sets.  

\subsection{GEH Distance Function}
\label{sec:geh_distance_function}
We define a function for computing the distance between two histograms 
$h_m^A$ and $h_m^B$ associated with the same traffic movement $m$ and scenarios $A$ and $B$. The function is based on the GEH statistic, which is widely used in traffic engineering to compare vehicular flows.

Given two vehicle flows $A$ and $B$, expressed in vehicles per hour, the GEH statistic is computed as 
\begin{equation}
    GEH(A, B) = \sqrt{\frac{2\,(A - B)^2}{A + B}}
    \label{eq:geh}
\end{equation}
The GEH statistic is often used thanks to its property of being scale-insensitive for the typical range of traffic flows, as it provides a consistent acceptance threshold across both high- and low-volume links, unlike percentage error measures that over-penalize small flows.  In practice, an acceptance threshold GEH $< 5$ is used \cite{fhwaTAT2004}.

Our histogram distance measure $d$ between two histograms $h_m^A$ and $h_m^B$ (for the same traffic movement $m$ and scenarios $A$ and $B$) is therefore defined as:
\begin{equation}
\label{eq:geh_dist}
    d(h_m^A, h_m^B) 
    = \sum_{k=1}^{K} 
    \mathbbm{1}\left(
    \sqrt{\frac{(h_m^A(k) - h_m^B(k))^2}
    {h_m^A(k) + h_m^B(k)}} > 5 \right).
\end{equation}
where $\mathbbm{1}(\cdot)$ is the indicator function. 

More informally, our functions compares the histograms bucket-by-bucket, computes the GEH distance between the two values for the same bucket, and counts how many have a GEH distance higher than 5 (the standard acceptance threshold in traffic engineering).

\section{Experimental Setup}

We evaluate the proposed distance measures in a controlled simulation study using the simulation tool, Simulation of Urban MObility (SUMO). SUMO is a microscopic traffic simulator that allows to simulate traffic flows and signal control policies with high precision (sub-second simulation steps) and to record key metrics such as throughput and travel time. 

The objective of our experimental setup is to determine whether our histogram-based scenario distances correlates with changes in controller performance. In other words, we want to observe whether the change in performance that a controller trained/calibrated for a particular scenario $S$ experiences when evaluated in a scenario $S' \neq S$ is correlated to the measured distance between $S$ and $S'$, according to our method.

\begin{figure}
    \centering
    \includegraphics[width=\linewidth]{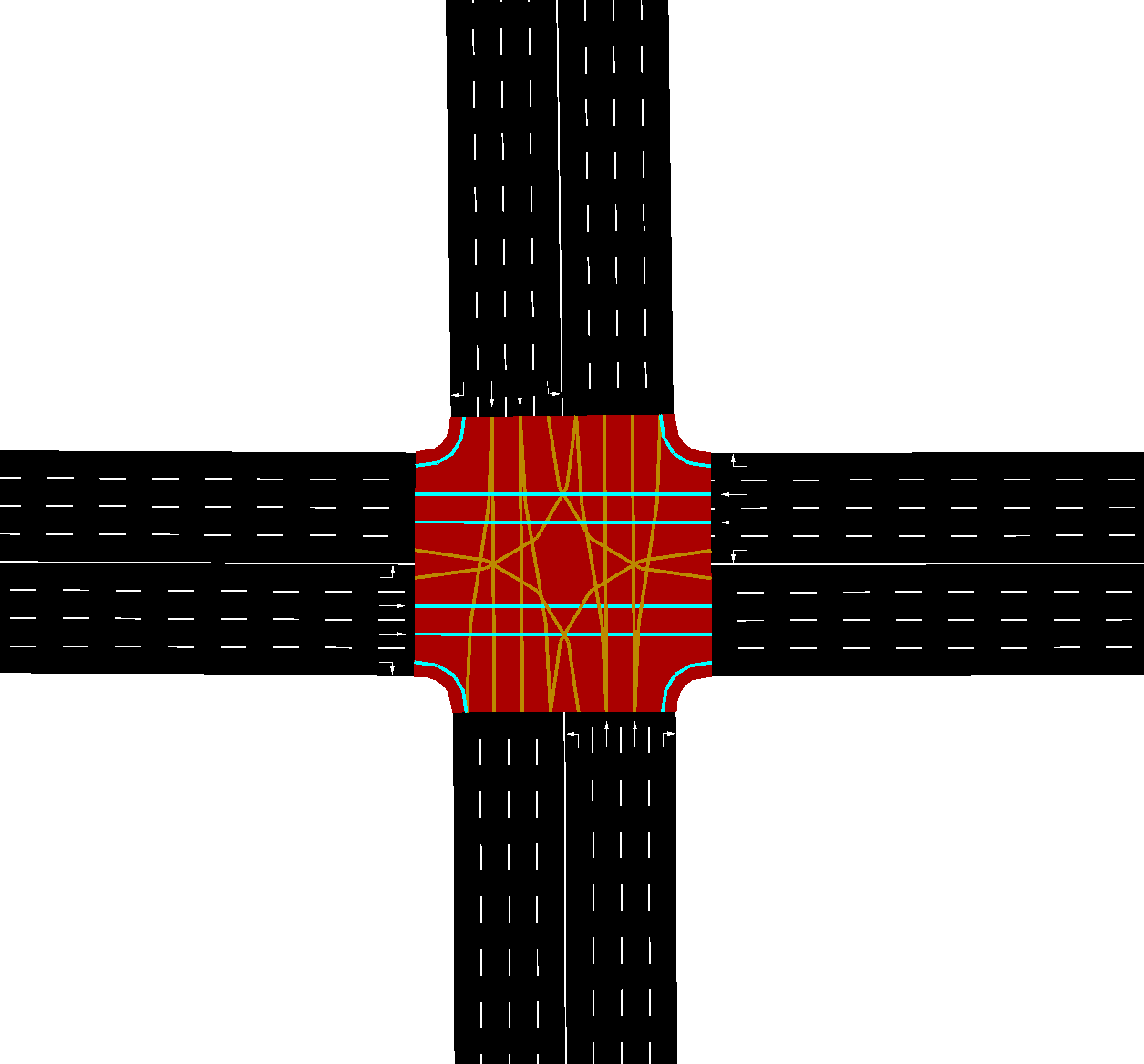}
    \caption{The simulated intersection with its traffic movements, as viewed in the \textit{netedit} tool of the SUMO software. Note that while SUMO represents each lane-to-lane collection as a separate movement, we group together all movements by their incoming and outgoing aproach. This means that in, this case, each approach has the three traffic movements shown in Figure \ref{fig:traffic_movements}}.
    \label{fig:intersection}
\end{figure}

Experiments are conducted in SUMO on a single four-leg intersection with realistic lane configurations and turning movements. Figure \ref{fig:intersection} shows the simulated intersection along with its traffic movements. Vehicles follow fixed routes determined by scenario demand distributions.  

\subsection{Scenario Generation}
We generate a set of 20 distinct traffic scenarios by varying the distribution of vehicles across traffic movements while keeping the overall demand level at 4000 vehicles. To generate a traffic scenario, we generate a set of 4000 random departure times in the range $[0, 3600]$ (seconds). Then, we randomly assign a traffic movement to each. We therefore generate a route file, required by SUMO, containing a specification of the vehicles that will be simulated, each with its departure time and traffic movement. This results in scenarios that range from balanced to highly skewed demand conditions.  

\subsection{Controllers}
Two controllers are used to evaluate performance across the generated scenarios:  
\begin{itemize}
    \item \textbf{NEMA-style actuated controller}, configured in free mode to reflect standard practice in an isolated intersection (no coordination with other intersections is active).  
    \item \textbf{Reinforcement Learning controller}, using the FRAP++ \cite{frap} algorithm.   
\end{itemize}

This combination allows us to investigate both rule-based and learning-based control strategies.  

\subsection{Performance Metrics}
We measure controller performance using two metrics:  
\begin{itemize}
    \item \textbf{Average travel time} of vehicles from when they are generated by the simulator to when they manage to cross the stopbar at the intersection.  
    \item \textbf{Throughput}, defined as the number of vehicles crossing the stopbar during the simulation horizon.  
\end{itemize}

\subsection{Evaluation Procedure}
For each scenario $S$ in the set of 20 scenarios, we train or calibrate a controller (NEMA or FRAP) specifically on $S$. We then evaluate the resulting controller in all 20 scenarios, including $S$ itself. Repeating this procedure for every scenario produces $20 \times 20 = 400$ evaluations per controller type.  

During each evaluation, we record the performance metrics defined in the previous subsection (average travel time and throughput). Finally, we analyze the correlation between scenario-to-scenario distances and the corresponding degradation in controller performance across scenarios.

\section{Results}

\begin{figure*}[h!]
    \centering
    \includegraphics[width=\linewidth]{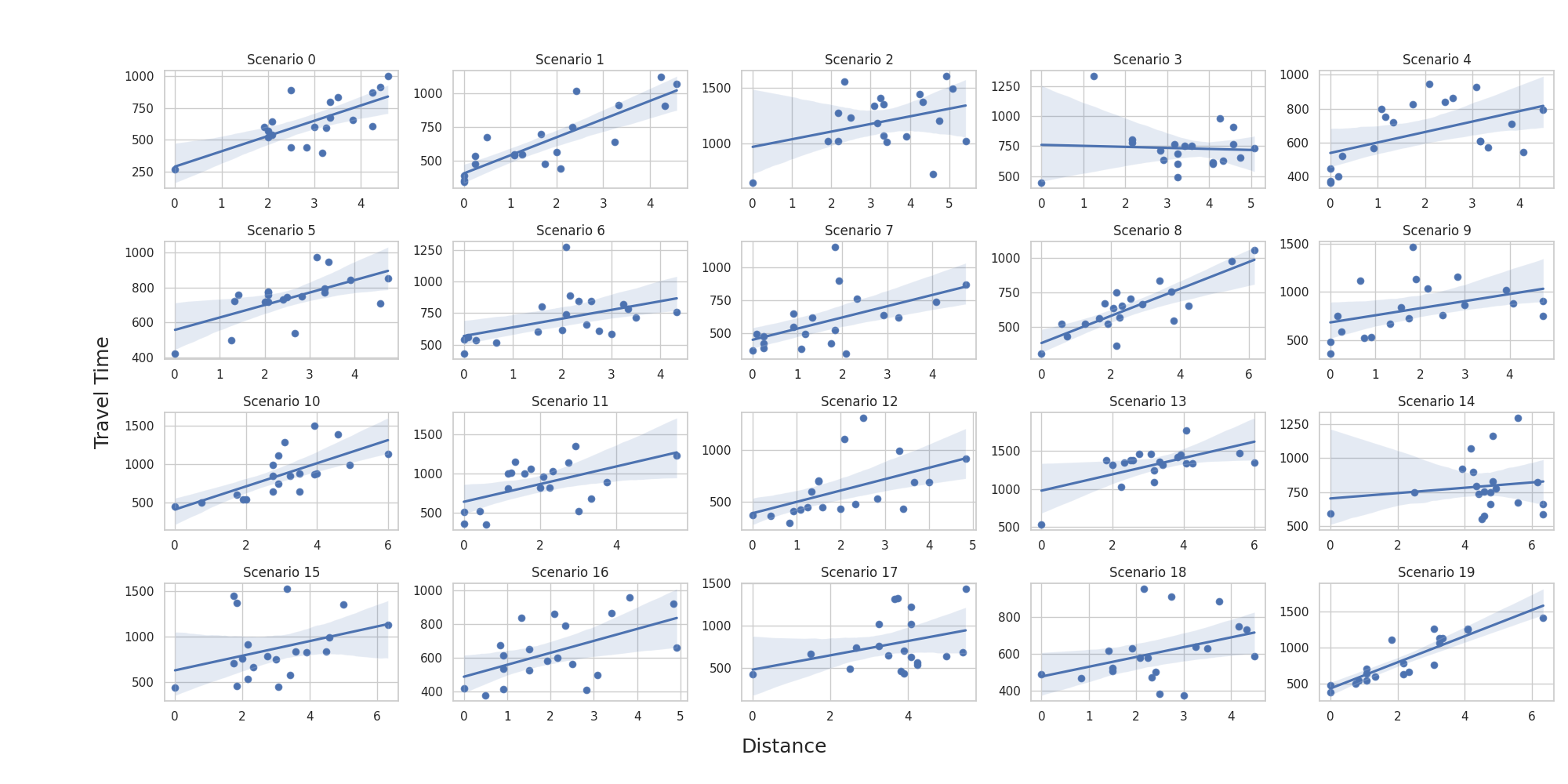}
    \caption{Travel time as a function of GEH distance for the 20 FRAP++ models}
    \label{fig:frap_multi_plot_travel_time}
\end{figure*}

\begin{figure*}[h!]
    \centering
    \includegraphics[width=\linewidth]{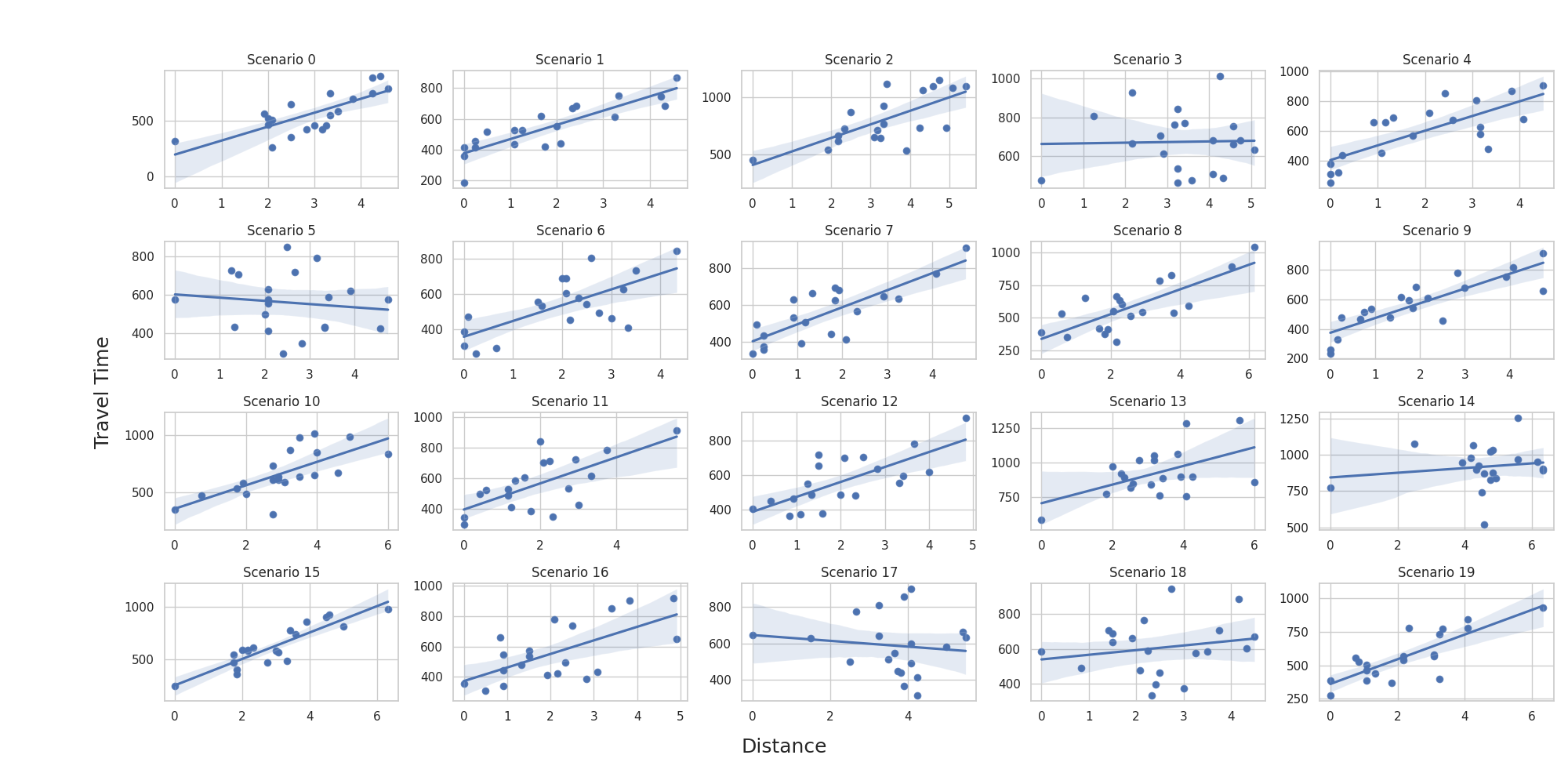}
    \caption{Travel time as a function of GEH distance for the 20 NEMA models}
    \label{fig:nema_multi_plot_travel_time}
\end{figure*}

We evaluate the predictive power of the proposed distance function by analyzing its relationship with performance degradation across scenarios. Figures~\ref{fig:frap_multi_plot_travel_time} and~\ref{fig:nema_multi_plot_travel_time} show the scatter plots of average travel time against scenario distance for FRAP++ and NEMA controllers, respectively, and the corresponding linear regression fitted to the data-points with its 95\% confidence interval. In both cases, larger distances correspond to higher travel times, indicating that the metric captures meaningful shifts in scenario difficulty.  

% --------------- MULTI REGRESSION PLOTS -----------------------

\begin{figure*}[h!]
    \centering
    \begin{subfigure}[b]{0.49\textwidth}
        \includegraphics[width=\linewidth]{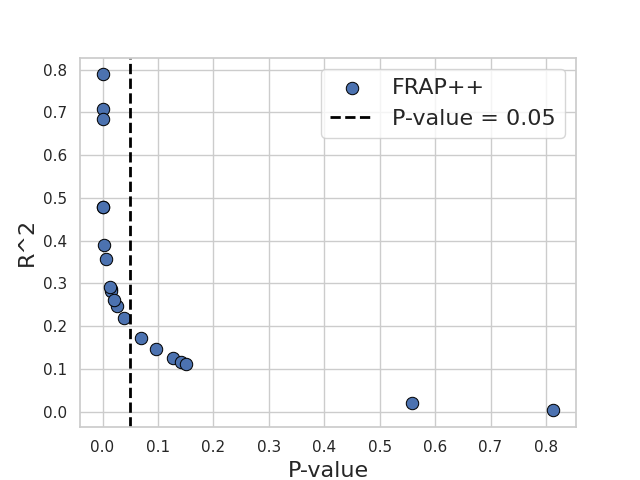}
        \caption{FRAP++ models}
        \label{fig:frap_distance_travel_time_p_r}
    \end{subfigure}
    \hfill
    \begin{subfigure}[b]{0.49\textwidth}
        \includegraphics[width=\linewidth]{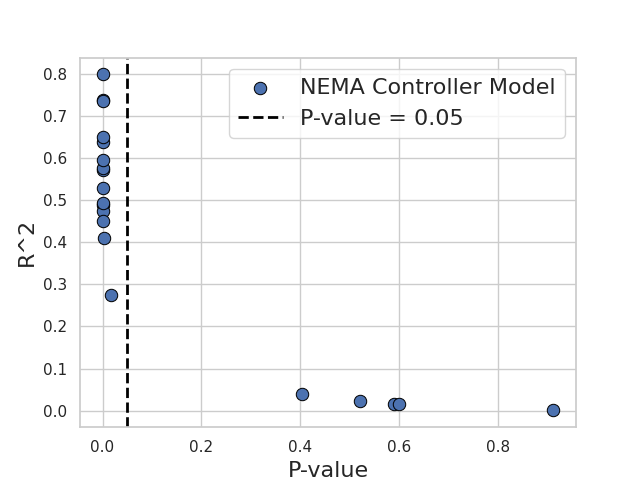}
        \caption{NEMA models}
        \label{fig:nema_distance_travel_time_p_r}
    \end{subfigure}
    \caption[$R^2$ agains P value for GEH-Distance-with-Threshold-C]{$R^2$ against P value for regressions of Figure \ref{fig:frap_multi_plot_travel_time} and Figure \ref{fig:nema_multi_plot_travel_time} for \textbf{GEH distance}}
    \label{fig:distance_travel_time_p_r}
\end{figure*}

\begin{figure*}[h!]
    \centering
    \begin{subfigure}[b]{0.49\textwidth}
        \includegraphics[width=\linewidth]{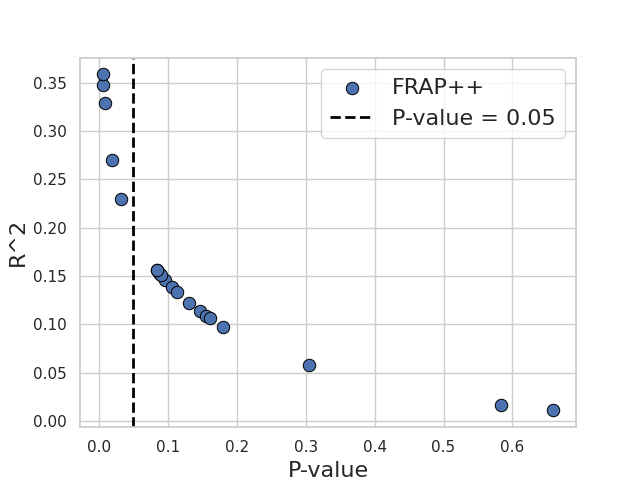}
        \caption{FRAP++ models}
        \label{fig:frap_kl_distance_travel_time_p_r}
    \end{subfigure}
    \hfill
    \begin{subfigure}[b]{0.49\textwidth}
        \includegraphics[width=\linewidth]{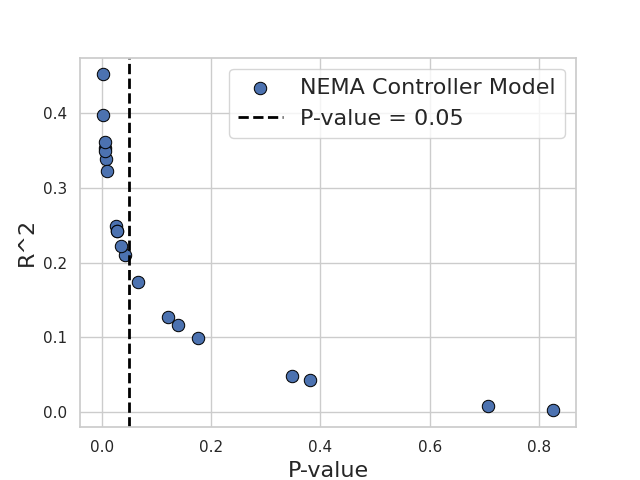}
        \caption{NEMA models}
        \label{fig:nema_kl_distance_travel_time_p_r}
    \end{subfigure}
    \caption[$R^2$ agains P value for KL Distance regressions]{$R^2$ against P value for regressions for the approach of \cite{tonguz2025kl} based on KL Distance }
    \label{fig:distance_kl_travel_time_p_r}
\end{figure*}

To assess robustness across all training scenarios, Figure~\ref{fig:distance_travel_time_p_r} reports the $R^2$ and $p$-values of linear regressions fitted for each controller. Here, the $R^2$ coefficient measures how much of the variation in travel time can be explained by the distance metric: higher values indicate a stronger linear relationship. The $p$-value tests whether this relationship is statistically significant, with $p < 0.05$ indicating that the correlation is unlikely to be due to random chance. The majority of regressions are statistically significant ($p < 0.05$), with FRAP++ generally achieving stronger correlations than NEMA. This suggests that the GEH distance with threshold is a reliable predictor of performance degradation, particularly for learning-based controllers.  For comparison, Figure\ref{fig:distance_kl_travel_time_p_r} shows the same statistics for the approach of \cite{tonguz2025kl}, which is based on KL distance computed on hourly vehicle volumes --while our histogram-based approach looks at multiple time intervals of 5 minutes each. It is clear how, while in our approach most data points stay on the left of the 0.05 p-value line and have an high $R^2$ coefficient, in the KL-based approach most data points lie on the right side of the 0.05 p-value line and $R^2$ coefficients are much lower. This indicates that our proposed approach is substantially more effective in predicting performance degradation under distribution shift.

% ---------------------------- P VALUE ANALYSIS ------------------------

\begin{figure*}
    \centering
    \begin{subfigure}[b]{0.49\linewidth}
    \includegraphics[width=\linewidth]{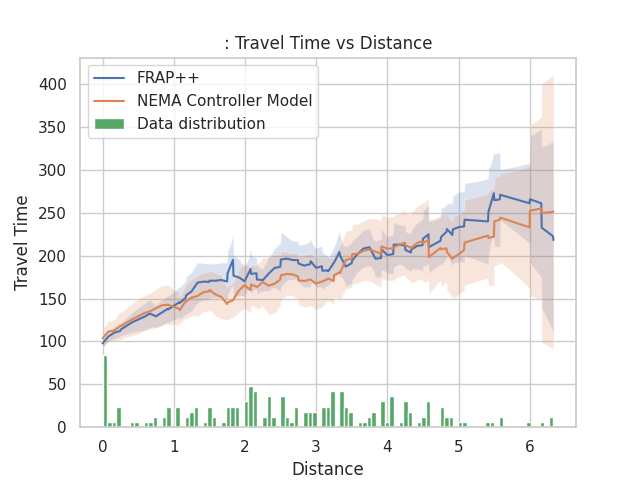}
    \caption{Travel time}
    \label{fig:res_geh_threshold_averaged_travel_time}
    \end{subfigure}
    \hfill
        \begin{subfigure}[b]{0.49\linewidth}
    \includegraphics[width=\linewidth]{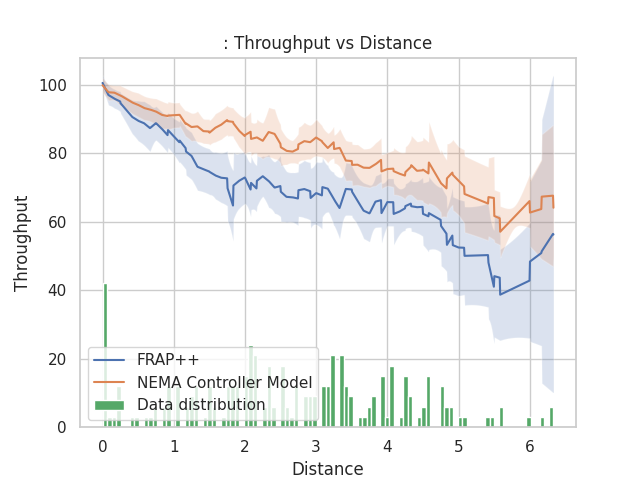}
    \caption{Throughput}
    \label{fig:res_geh_threshold_averaged_throughput}
    \end{subfigure}
    \hfill
    \caption{Averaged travel time and throughput versus GEH distance}
    \label{fig:res_geh_threshold_averaged}
\end{figure*}

Figure~\ref{fig:res_geh_threshold_averaged} aggregates results across all 20 scenarios, plotting both average travel time and throughput against the GEH distance. We observe a clear positive trend between distance and travel time as well as a negative trend with throughput. These results confirm that the proposed metric not only captures the distribution shift but also predicts its impact on key performance indicators.

% ---------------------- AVERAGED PLOTS -----------------

\section{Discussion}

The results demonstrate that our approach can be successfully used to measure the distribution shift in traffic scenarios. In particular, we observe that higher scenario distances, as measured by our GEH-based distance, consistently correspond to higher levels of performance degradation, both in terms of increased travel time and reduced throughput. This confirms that the metric captures the kinds of demand change that most affect the effectiveness of traffic signal controllers.  

An important property of the GEH-based distance is its interpretability. Because GEH is already widely used in transportation engineering, practitioners can relate distance values to familiar thresholds. The use of a cutoff value ($GEH < 5$) further simplifies interpretation: the metric effectively counts how many time histogram intervals differ "substantially" between two scenarios. This representation provides 
a bridge between domain knowledge in traffic engineering and the needs of algorithmic evaluation.  

The results also suggest differences between controllers. FRAP++, the learning-based controller, shows stronger correlations between distance and performance degradation compared to NEMA. This may be explained by the fact that rule-based controllers are tuned to broad traffic patterns, making them somewhat robust to distribution shifts, whereas data-driven controllers are optimized for a specific training distribution and degrade more sharply when conditions change.  

Despite its advantages, the GEH distance with threshold has limitations. By reducing the differences to a count of “mismatched” bins, the measure discards information about the magnitude of deviations once the threshold is exceeded. Moreover, despite a visible correlation trend, results do not show perfect correlations, with some scenarios exhibiting a lower degree of alignment between distance values and observed performance degradation. This suggests that while the proposed metric captures the main effect of distribution shift, additional factors remain unaccounted for.  

Our approach is straightforward to implement if one has access to traffic counts for each traffic movement at signalized intersections, which are often collected by sensors such as cameras, radars, or induction loops. Therefore, our method can be immediately implemented in traffic management systems that have such sensors, and could be used to detect when traffic flows are deviating from the norm, indicating that the data used to train or fine-tuned the control system might have become outdated.  

In general, the findings of our study show that our method can serve as a practical tool for quantifying distribution shift in traffic control research. Furthermore, the approach can support systematic scenario design and benchmarking of algorithms under controlled shifts, ensuring that new controllers are tested under diverse and representative conditions. In operational settings, such metrics could guide the selection of training scenarios, enable adaptive retraining of reinforcement learning controllers, and provide monitoring tools to detect when deployed systems encounter traffic demand that deviates from historical patterns. More broadly, the method contributes to building robust and explainable 
evaluation pipelines, helping bridge the gap between research prototypes and field-ready traffic management solutions.

\section{Conclusions}

This paper introduces a framework for quantifying the distribution shift in traffic  signal control based on histograms at the traffic movement level and a GEH-based distance function with thresholding. The approach provides a policy-independent representation of traffic demand and a domain-specific measure of dissimilarity between scenarios. Experimental results on a set of 20 simulated scenarios show that larger distances consistently correspond to degraded controller performance, capturing increases in travel time and decreases in throughput. The method proves particularly effective in explaining the sensitivity of learning-based controllers such as FRAP++, while also providing meaningful insights for rule-based controllers like NEMA.  

Beyond its empirical performance, the proposed approach offers interpretability and practical relevance. Histograms align with how traffic engineers reason about movements through an intersection, while the GEH statistic is already established in transportation engineering as a standard validation tool. Their combination produces a metric that is intuitive, reproducible, and easily communicated to both researchers and practitioners.  

The broader implications are significant. By enabling more robust evaluation of traffic control algorithms under diverse conditions, this work supports the development of controllers that are more reliable, sustainable, and fair. Improved robustness can translate into reduced congestion, fuel use, and emissions, benefiting both urban mobility and environmental sustainability. In addition, transparent and reproducible evaluation metrics contribute to more trustworthy research practices, helping to prevent bias in the way algorithms are benchmarked and compared.  

While the results are promising, several possible directions remain for future work. These include extending the approach to multi-intersection networks, refining the thresholding mechanism to capture both the extent and the magnitude of deviations, and validating the method with real-world traffic data. Such developments would further strengthen the role of distribution-shift metrics in bridging the gap between algorithmic innovation and field-ready traffic management systems.

\FloatBarrier

\balance

\end{document}